\documentclass{article}

\usepackage{graphicx}

\begin{document}

\title{Electron Transmission through Modified Benzene}

\date{July 5, 2016}

\author{Dong Qiu\thanks{School of Mathematical  and Computational Sciences, University of Prince Edward Island, Charlottetown, PE, C1A 4P3, Canada} \thanks{current address: Department of Mathematics and Statistics, Concordia University, Montreal, QC, H3G 1M8, Canada}\and Kenneth W. Sulston\footnotemark[1] \thanks{corresponding author (sulston@upei.ca)}}

\maketitle

\section{Abstract}

The renormalization method is applied to investigate the electron transmission properties of a circuit
containing a benzene molecule, in which one of the carbon atoms has been modified so as to simulate
displacement in position or replacement by another atom.
Consideration of the different possible attachments of the leads, and the relative location of the modified
atom, results in 9 distinct configurations to examine.
For each configuration, the number and locations of anti-resonances, and whether they shift upon
variation of the parameters, is seen to be the key to determining the shape of the electron-transmission
curve. 
In particular, those configurations, in which the perturbed atom is not directly attached to a lead,
are seen to have the most variation in their structure, compared to pure benzene.

\section{Introduction}

In this era of nanotechnology, it is becoming increasingly feasible to construct molecular wires and circuits,
with specific constituents, and in desired configurations \cite{ref1}. 
Thus, it is important and interesting to study electron transport through fundamental structures.
One of the most common organic molecules is benzene, which is important due to its structural
simplicity and its appearance in larger compounds.
Consequently, in this article, we will be extending previous work, studying electron transmission
through an atomic wire containing a single benzene molecule.
Sulston and Davison \cite{ref2} used the renormalization technique \cite{ref3} to rework the benzene
molecule as a dimer with rescaled parameters, connected to leads in each of the three possible
configurations, namely para, meta and ortho.
(More generally, they also used the method to consider 2 or 3 benzenes connected in series or parallel,
although here we discuss circuits of only a single benzene molecule.)
They then used the Lippmann-Schwinger equation to calculate the transmission-energy probability function $T(E)$
for each case. 
The 3 circuit types yielded very distinctive graphs, with the transmittivities largely determined by the number
and positions of resonances and anti-resonances, with a general conclusion being that para-benzene is the 
strongest transmitter, and meta-benzene the weakest.

In this paper, we extend the work of Sulston and Davison, by considering a modified benzene molecule,
wherein one of the carbon atoms is replaced by another atom (or, alternatively, a carbon atom is
displaced from its normal position).
The methodology is that of \cite{ref2}, where the modification to the molecule is modelled by changing
the site energy for the perturbed atomic site, and its bond energies with the two adjacent carbon
sites.
The perturbed atom can have different locations relative to the sites attached to the leads, so that there
are a fairly large number of possible configurations, resulting in a diversity of behaviours for the $T(E)$ curves.

\section{Theory}

\begin{figure}[h]
\includegraphics[width=12cm]{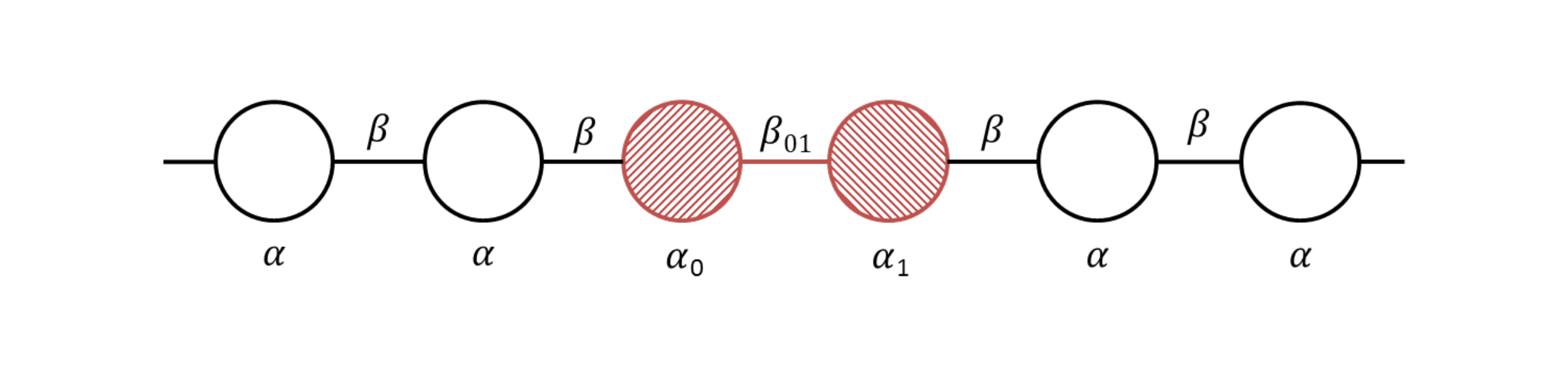}
\caption{Atomic wire containing asymmetric dimer impurity.}
\label{fig1}
\end{figure}

The methodology utilized the Lippmann-Schwinger theory \cite{ref4,ref5} applied to electron transport through an
atomic wire containing an asymmetric dimer impurity, as shown in Figure \ref{fig1}.
The wire itself consists of an infinite chain of atoms, with site energy $\alpha$ on each atom, and bond energy
$\beta$ between neighbouring atoms. 
Embedded within it is an asymmetric dimer impurity occupying two adjacent sites, with the properties that
the site energies are changed to $\alpha_0 = \alpha +\varepsilon_0$ and $\alpha_1 = \alpha +\varepsilon_1$, and
the bond energy between these sites is modified to $\beta_{01} = \beta + \Delta\beta$.
The electron-transmission properties of this structure have been well-studied \cite{ref2,ref6}, and
the transmission-energy probability function $T(E)$ has the form
\begin{equation}
T(E)= {{(1+2\gamma)^2 (4-X^2)} \over {(1-2Q)^2 (4-X^2) + 4(P-QX)^2}}  ,
\label{eq1}
\end{equation}
where
\begin{equation}
P = z_0+z_1  ~,~ Q = z_0 z_1 - \gamma -\gamma^2 ,
\label{eq2}
\end{equation}
with
\begin{equation}
z_{0,1} = ({\alpha}_{0,1} - \alpha)/ 2\beta ~,~ \gamma = ({\beta}_{01} -\beta)/ 2\beta ,
\label{eq3}
\end{equation}
and the reduced dimensionless energy is
\begin{equation}
X= (E - \alpha)/\beta .
\label{eq4}
\end{equation}
This model provides a very efficient framework for studying transmission through a wide variety of systems \cite{ref3},
provided that the system can be reduced to a dimer via the renormalization method (as discussed in the next section),
with rescaled parameters $\alpha_0$, $\alpha_1$ and $\beta_{01}$.
The transmission function is then calculated by way of (\ref{eq1}).

\section{Renormalization Method}

\begin{figure}[h]
\includegraphics[width=12cm]{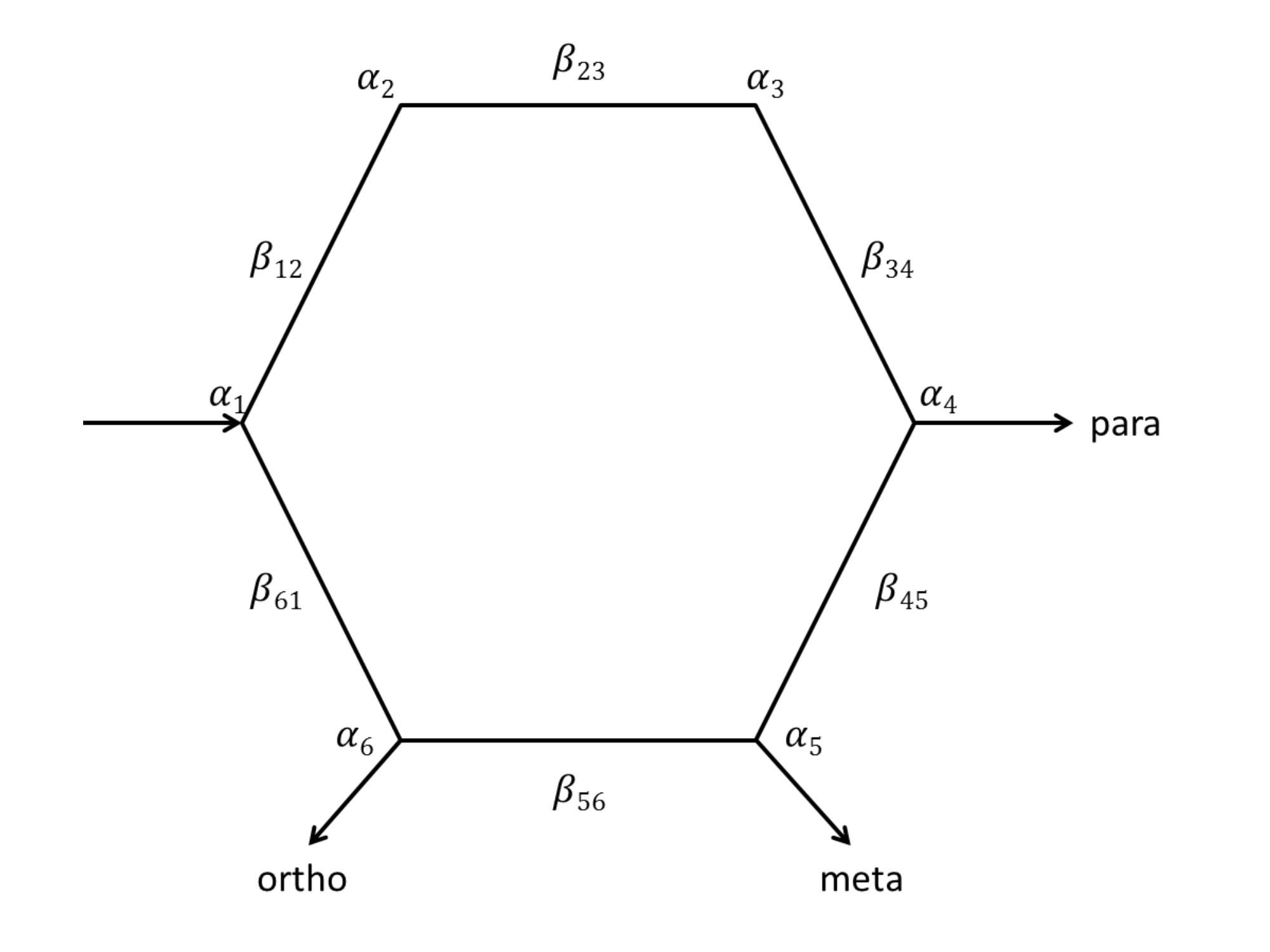}
\caption{Modified benzene molecule, showing para, meta and ortho connections to leads.}
\label{fig2}
\end{figure}

The model under consideration here is shown in Figure \ref{fig2}, consisting of a modified benzene ring attached
to atomic wire leads.
The ring consists of 6 atoms, with site energies $\alpha_n$ (for $n=1, \dots, 6$), with adjacent atoms $n$ and $n+1$ connected
by a bond of energy $\beta_{n,n+1}$.
In ordinary benzene, the atoms are all identical, so that $\alpha_n = \alpha$ and $\beta_{n,n+1} = \beta$.
For the systems being considered here, the benzene is modified so that one of the atoms (say, $l$) is distinct from the others,
which is modelled by taking $\alpha_l  = \bar{\alpha} \ne \alpha$ and $\beta_{l-1,l}=\beta_{l,l+1} = \bar{\beta} \ne \beta$.
Referring again to Figure \ref{fig2}, the three possible connection schemes for the leads are: (i) para, in which the leads are
attached at opposite ends of the benzene, i.e. positions 1 and 4, (ii) meta, where they are attached at positions 1 and 5,
and (iii) ortho, using adjacent positions, namely 1 and 6.
With this in mind, there would seem to be 18 possible combinations to consider, specifically,
modification of each of atoms 1 through 6 for each of the 3 types of leads. 
However, due to symmetry considerations, there are actually fewer unique cases.
For example, in the para case, modifying atom 1 is equivalent to modifiying atom 4, while modifications
to atoms 2, 3, 5 and 6 are equivalent.
Thus, in the para case, we only need to consider modifications to atoms 1 and 2, which we denote as
para-1 and para-2, respectively.
By similar reasoning, the remaining unique combinations are meta-1, meta-2, meta-3, meta-6, ortho-1, ortho-2
and ortho-3, making a total of 9 cases to consider.

\begin{figure}[hp]
\includegraphics[width=7cm]{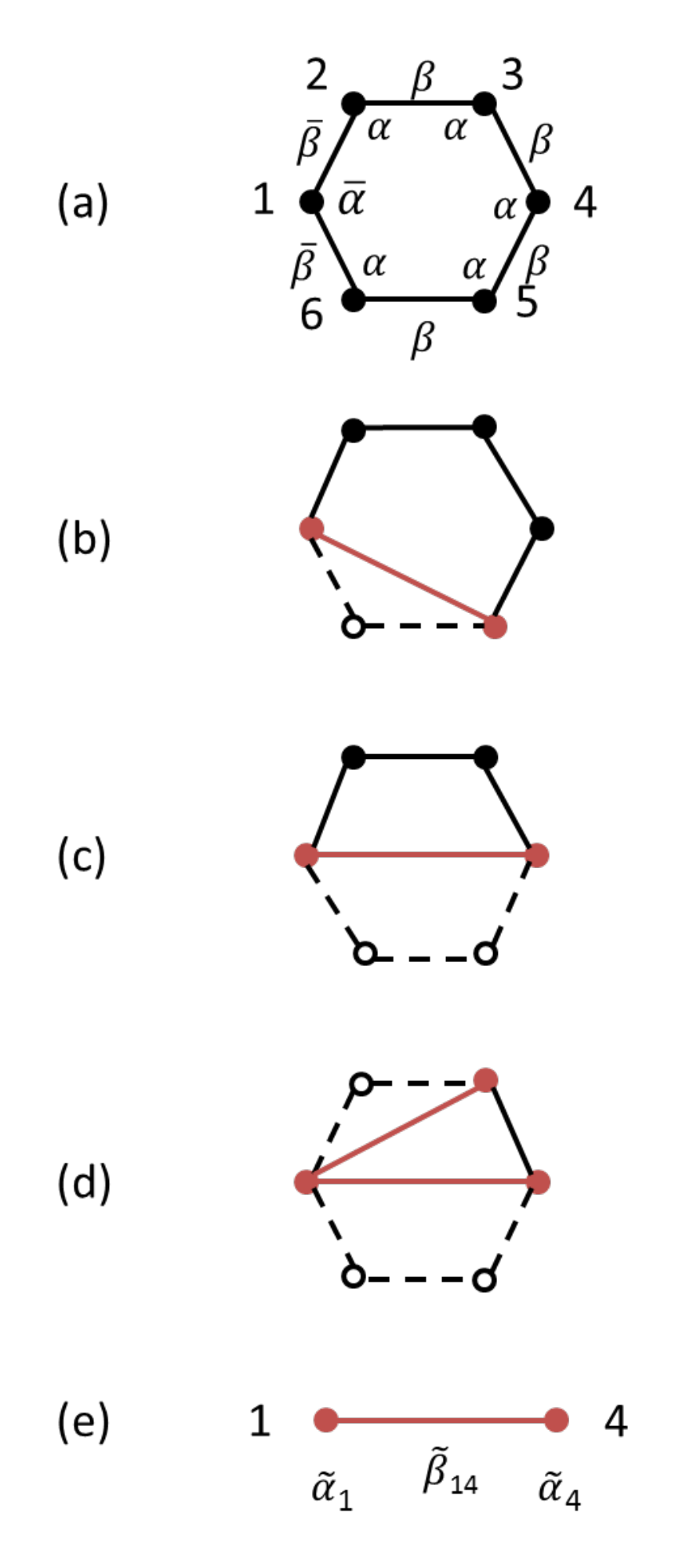}
\caption{Renormalization process for reducing the para-1 benzene molecule to a dimer. (a) shows original molecule,
while (b)-(e) show decimation of sites 6, 5, 2, 3, respectively, with final dimer in (e). The original sites (bonds)
are shown as solid black dots (lines). Rescaled sites (bonds) are shown as solid red dots (lines). Decimated sites (bonds)
are shown as open black circles (dashed black lines).}
\label{fig3}
\end{figure}

Each of the 9 modified benzene molecules can be reduced to a dimer, by means of the renormalization method \cite{ref3},
which eliminates, one-by-one, the 4 atoms of the ring not attached to leads, leaving the remaining two atoms, with rescaled
parameters, attached to the rest of the wire. 
The derivation of the method is given in \cite{ref2}, for example, but the key result is that elimination of atom $n$
leaves its nearest-neighbour atoms $n \pm 1$, with rescaled parameters given by
\begin{equation}
\tilde{\alpha}_{n-1} =\alpha_{n-1}+ { {\beta}_{n-1,n}^2 \over {E-\alpha_n}},
\label{eq5}
\end{equation}
\begin{equation}
\tilde{\alpha}_{n+1} =\alpha_{n+1}+ { {\beta}_{n,n+1}^2 \over {E-\alpha_n}},
\label{eq6}
\end{equation}
\begin{equation}
\tilde{\beta}_{n-1,n+1}  = {{\beta_{n-1,n} \beta_{n,n+1}} \over {E-\alpha_n}}.
\label{eq7}
\end{equation}
The renormalization process is shown schematically in Figure \ref{fig3}, using para-1 benzene as an example.
The renormalization equations (\ref{eq5})-(\ref{eq7}) are applied at each step, and the final rescaled 
parameters for the dimer can then be calculated recursively. 
For the sake of brevity, we give below only the final rescaled parameters for each of the 9 cases, noting
again that for each case, the dimer consists of the rescaled sites originally attached to the leads.
For convenience, the results are expressed using reduced energies $X$ from  (\ref{eq4}) and
\begin{equation}
Y = (E-\bar{\alpha})/\bar{\beta}.
\label{eq8}
\end{equation}
\newpage 
\noindent {\large {\bf para-1}}

\vspace{10pt}

\begin{equation}
\tilde{\alpha}_1 =\bar{\alpha}+ { 2\bar{\beta}^2 X \over {\beta (X^2-1)}},
\label{eq9}
\end{equation}
\begin{equation}
\tilde{\alpha}_4 =\alpha+  {2 {\beta X} \over { X^2-1}},
\label{eq10}
\end{equation}
\begin{equation}
\tilde{\beta}_{14}  =  {2\bar{\beta} \over { X^2-1}}.
\label{eq11}
\end{equation}

\noindent {\large {\bf para-2}}

\vspace{10pt}

\begin{equation}
\tilde{\alpha}_1 =\alpha + { \beta X \over {X^2-1}} + { \beta {\bar{\beta} X} \over {\beta XY-\bar{\beta}}},
\label{eq12}
\end{equation}
\begin{equation}
\tilde{\alpha}_4 = \alpha + { \beta X \over {X^2-1}} + {{ \beta^2 Y} \over {\beta XY-\bar{\beta}}},
\label{eq13}
\end{equation}
\begin{equation}
\tilde{\beta}_{14}  =  {{\beta} \over { X^2-1}} + {{\beta \bar{\beta}} \over {\beta XY- \bar{\beta}}}.
\label{eq14}
\end{equation}

\noindent {\large {\bf meta-1}}

\vspace{10pt}

\begin{equation}
\tilde{\alpha}_1 =\bar{\alpha}+ {\bar{\beta}^2 \over {\beta X}} \cdot {{2X^2-3} \over {X^2-2}},
\label{eq15}
\end{equation}
\begin{equation}
\tilde{\alpha}_5 =\alpha+  {\beta \over  X} \cdot {{2X^2-3} \over {X^2-2}},
\label{eq16}
\end{equation}
\begin{equation}
\tilde{\beta}_{15}  =  {\bar{\beta} \over X} \cdot { {X^2-1} \over {X^2-2}}.
\label{eq17}
\end{equation}

\noindent {\large {\bf meta-2}}

\vspace{10pt}

\begin{equation}
\tilde{\alpha}_1 = \alpha+ {\beta \over X} + {\beta \bar{\beta} (X^2-1) \over {X(\beta XY - \bar{\beta}) -\beta Y}},
\label{eq18}
\end{equation}
\begin{equation}
\tilde{\alpha}_5 = \alpha+ {\beta \over X} + {\beta (\beta XY -\bar{\beta}) \over {X(\beta XY - \bar{\beta}) -\beta Y}},
\label{eq19}
\end{equation}
\begin{equation}
\tilde{\beta}_{15} = {\beta \over X} + {\beta \bar{\beta} \over {X(\beta XY - \bar{\beta}) -\beta Y}} .
\label{eq20}
\end{equation}
\newpage 
\noindent {\large {\bf meta-3}}

\vspace{10pt}

\begin{equation}
\tilde{\alpha}_1 = \alpha+ {\beta \over X} \cdot {2\beta XY - 3\bar{\beta} \over {\beta XY -2\bar{\beta}}},
\label{eq21}
\end{equation}
\begin{equation}
\tilde{\alpha}_5 = \alpha+ {\beta \over X} \cdot {2 \beta XY - 3\bar{\beta} \over {\beta XY -2\bar{\beta}}},
\label{eq22}
\end{equation}
\begin{equation}
\tilde{\beta}_{15} = {\beta \over X} \cdot {\beta XY - \bar{\beta} \over {\beta XY -2\bar{\beta}}}.
\label{eq23}
\end{equation}

\noindent {\large {\bf meta-6}}

\vspace{10pt}

\begin{equation}
\tilde{\alpha}_1 = \alpha + {\bar{\beta} \over Y} + {\beta \over X} \cdot {X^2 -1 \over {X^2-2}} ,
\label{eq24}
\end{equation}
\begin{equation}
\tilde{\alpha}_5 = \alpha +{\bar{\beta} \over Y} + {\beta \over X} \cdot {X^2 -1 \over {X^2-2}} ,
\label{eq25}
\end{equation}
\begin{equation}
\tilde{\beta}_{15}  = {\bar{\beta} \over Y} + {\beta \over X} \cdot {1 \over {X^2-2}} .
\label{eq26}
\end{equation}

\noindent {\large {\bf ortho-1}}

\vspace{10pt}

\begin{equation}
\tilde{\alpha}_1 =\bar{\alpha}+ {\bar{\beta}^2 \over {\beta X}} +{\bar{\beta}^2 \over {\beta X}}  \cdot {X^2-1  \over {(X^2-1)^2-X^2}} ,
\label{eq27}
\end{equation}
\begin{equation}
\tilde{\alpha}_6 =\alpha+ {\beta \over X} +{\beta \over X} \cdot  {{ X^2-1} \over { (X^2-1)^2-X^2}},
\label{eq28}
\end{equation}
\begin{equation}
\tilde{\beta}_{16}  = \bar{\beta} + {\bar{\beta} \over {(X^2-1)^2-X^2 }}.
\label{eq29}
\end{equation}

\noindent {\large {\bf ortho-2}}

\vspace{10pt}

\begin{equation}
\tilde{\alpha}_1 =\bar{\alpha}+ {\bar{\beta} \over Y} + {\bar{\beta}^2  \over Y} \cdot {X^2 -1 \over  {(X^2-1)(\beta XY -\bar{\beta})- \beta XY  }},
\label{eq30}
\end{equation}
\begin{equation}
\tilde{\alpha}_6 =\alpha+ {\beta \over X} +  {\beta \over X}\cdot {\beta XY -\bar{\beta} \over {(X^2-1)(\beta XY -\bar{\beta})-\beta XY}},
\label{eq31}
\end{equation}
\begin{equation}
\tilde{\beta}_{16}  =  \beta + {\bar{\beta} \beta \over {(X^2-1)(\beta XY - \bar{\beta}) - \beta XY}}.
\label{eq32}
\end{equation}
\newpage 
\noindent {\large {\bf ortho-3}}

\vspace{10pt}

\begin{equation}
\tilde{\alpha}_1 =\alpha+ {\beta \over X} +{\beta \over X} \cdot {\bar{\beta} (X^2-1) \over {(X^2-1)(\beta XY -\bar{\beta})-\bar{\beta}X^2}},
\label{eq33}
\end{equation}
\begin{equation}
\tilde{\alpha}_6 =\alpha+ {\beta \over X} +{\beta \over X} \cdot {\beta XY -\bar{\beta} \over {(X^2-1)(\beta XY -\bar{\beta})-\bar{\beta}X^2}},
\label{eq34}
\end{equation}
\begin{equation}
\tilde{\beta}_{16}  = \beta +  {\beta \bar{\beta} \over {(X^2-1)(\beta XY -\bar{\beta})-\bar{\beta}X^2}}.
\label{eq35}
\end{equation}

\vspace{10pt}

\section{Results and Discussion}

It is now possible to evaluate $T(E)$ in (\ref{eq1}), for each of the 9 systems. For each case, one uses the corresponding trio of rescaled
parameters, given above, as the choice of $\alpha_0$, $\alpha_1$ and $\beta_{01}$, respectively, in (\ref{eq3}).
For parameter values, we chose $\alpha = 0$ and $\beta =-0.5$, which
leads to non-zero transmission probabilities for the energy range $-1 \le E \le 1$.
We vary $\bar{\alpha}$ and $\bar{\beta}$, to see the variety of behaviour that can arise in each case.

\begin{figure}[hp]
\includegraphics[width=12cm]{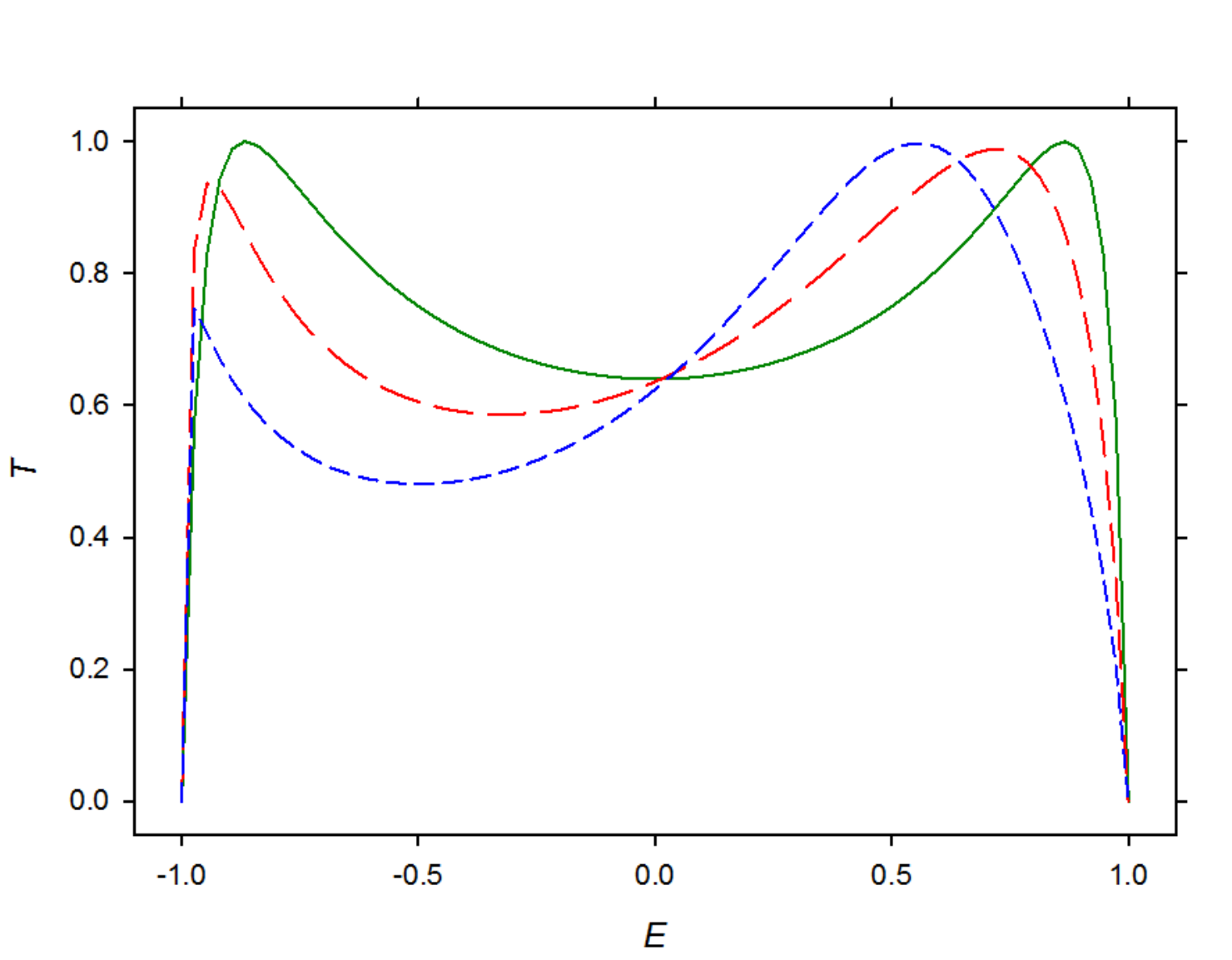}
\caption{Transmission $T$ versus energy $E$ for para-1 benzene, with $\bar{\beta}=-0.5$, and $\bar{\alpha}=$ (a) $0$ (green solid curve), (b) $-0.2$ (red long-dashed), (c) $-0.4$ (blue short-dashed).}
\label{fig4}
\end{figure}

Looking first at para-1 benzene, curves for the case where $\bar{\beta}=\beta =-0.5$ are shown in Figure \ref{fig4}. 
Figure \ref{fig4}(a) reproduces the previously published situation \cite{ref2} where $\bar{\alpha}=\alpha=0$,
corresponding to an unmodified (or ``pure'') para-benzene.
In this case, the $T(E)$ curve is symmetric about the band centre, with resonances at $E=\pm \sqrt3/2$,
but no anti-resonances.
Upon setting $\bar{\alpha} \ne \alpha$ (Figure \ref{fig4}(b,c)), the resonance peaks are shifted to lower energies,
in accord with the lowering of $\bar{\alpha}$, with the lower peak decreasing in height to non-resonance status.
The absence of anti-resonances persists.

\begin{figure}[hp]
\includegraphics[width=12cm]{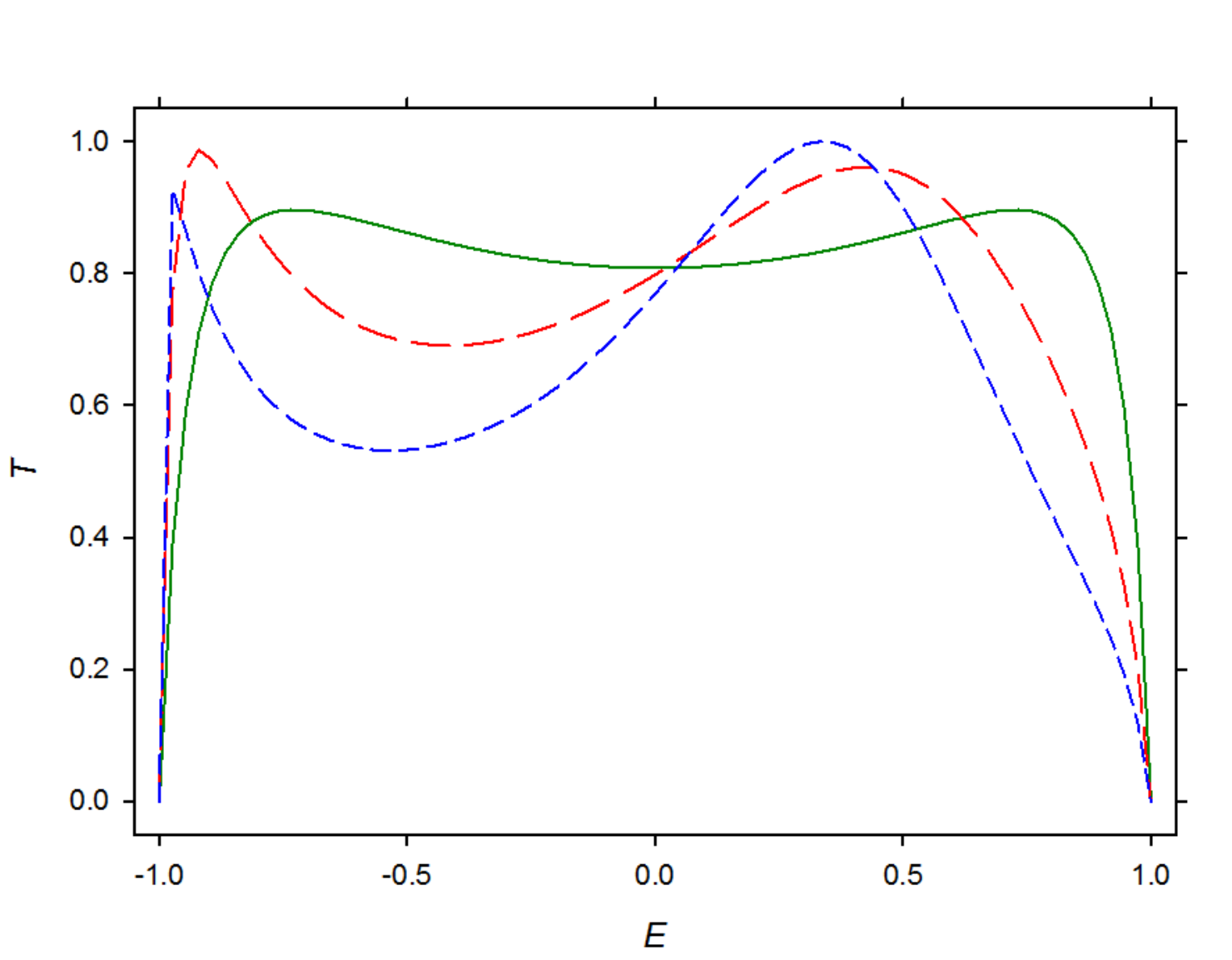}
\caption{Transmission $T$ versus energy $E$ for para-1 benzene, with $\bar{\beta}=-0.4$, and $\bar{\alpha}=$ (a) $0$ (green solid curve), (b) $-0.2$ (red long-dashed), (c) $-0.4$ (blue short-dashed).}
\label{fig5}
\end{figure}

\begin{figure}[hp]
\includegraphics[width=12cm]{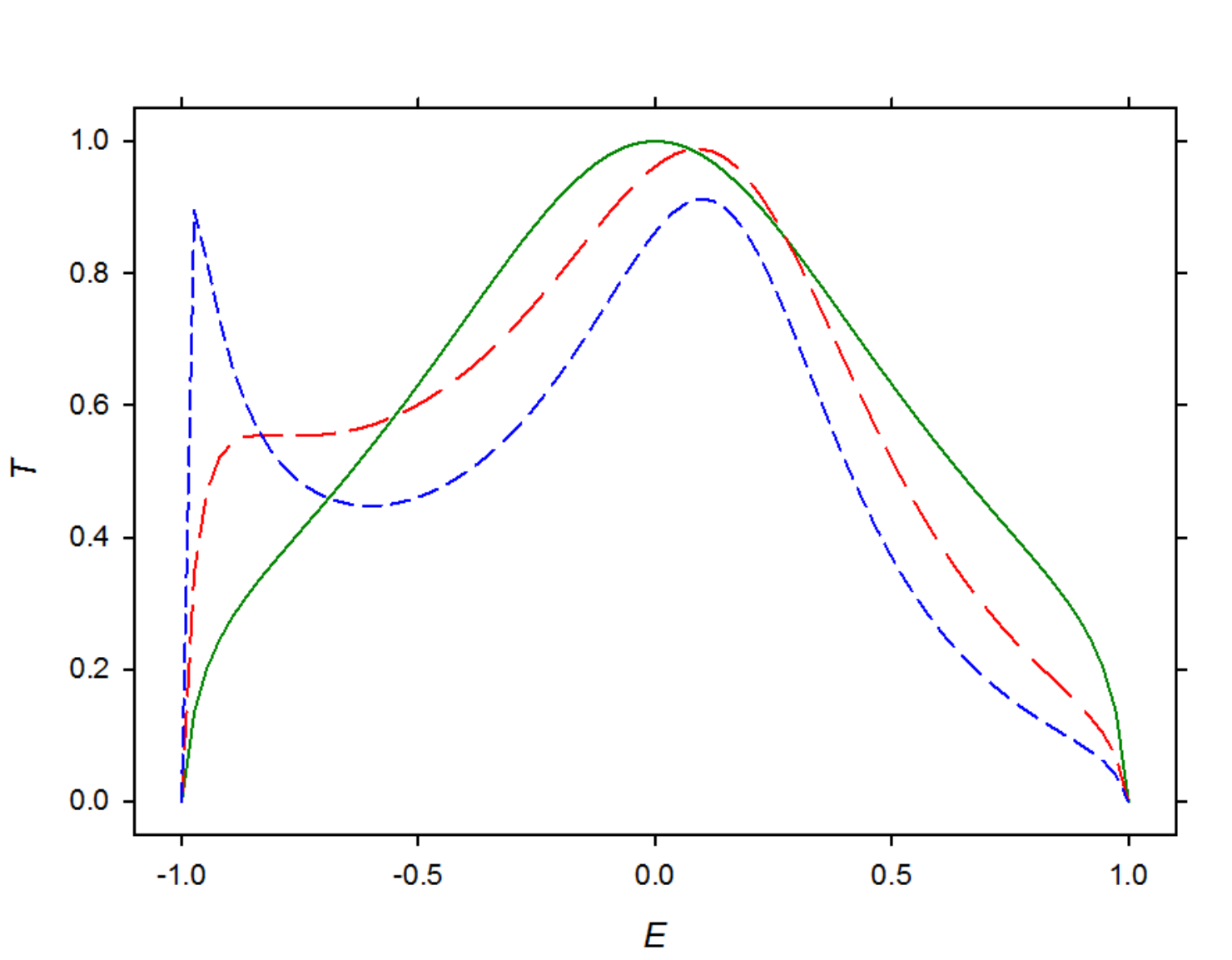}
\caption{Transmission $T$ versus energy $E$ for para-1 benzene, with $\bar{\beta}=-0.25$, and $\bar{\alpha}=$ (a) $0$ (green solid curve), (b) $-0.2$ (red long-dashed), (c) $-0.4$ (blue short-dashed).}
\label{fig6}
\end{figure}

Continuing with para-1, the curves for $\bar{\beta} =-0.4$ are shown in Figure \ref{fig5}.
The curves are similar to their counterparts in Figure \ref{fig4}, but with further rearrangement of the locations
and heights of the peaks, and specifically a tendency towards lower energies. 
This trend continues with further modification of $\bar{\beta}$ to a value of $-0.25$ (Figure \ref{fig6}), and in particular,
in graphs (a) and (b), the lower peak has been down-shifted far enough so as to assimilate it into the lower band edge.
Interestingly, in (a), resonance is restored in the upper peak.
In both Figures \ref{fig5} and \ref{fig6}, there are again no anti-resonances.
It might be considered that the value $\bar{\beta}=-0.25$ is somewhat extreme, and that the value of $-0.4$ in
Figure \ref{fig5} is more realistic, in which case we conclude that the effect of modifying $\bar{\beta}$ is fairly modest, in
comparison to that of $\bar{\alpha}$.
Also, it is important to note, here, and in fact in all the results, that the $T(E)$ curve is always symmetric whenever
$\bar{\alpha}=\alpha=0$ (the green solid curve (a) in each Figure).

\begin{figure}[hp]
\includegraphics[width=12cm]{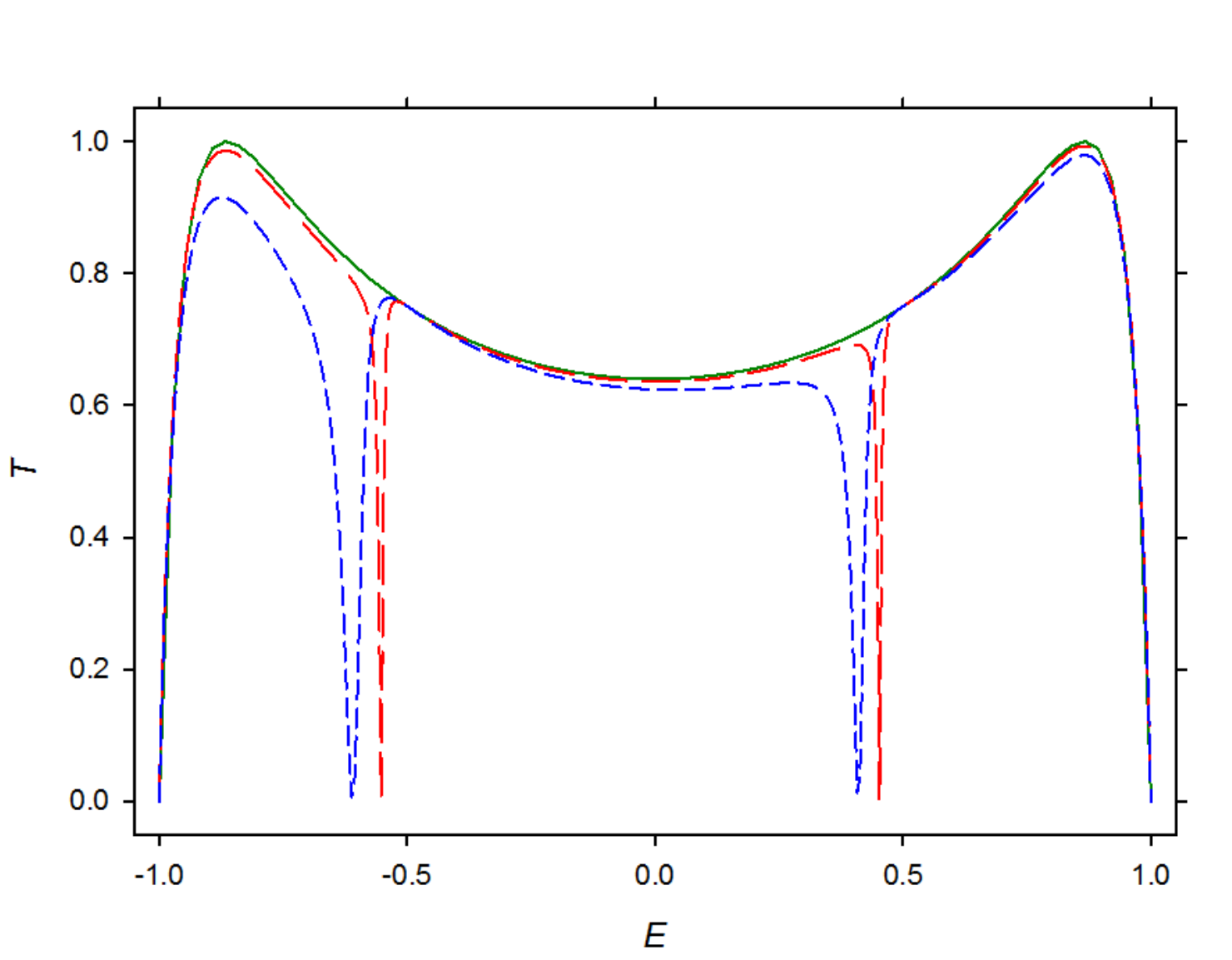}
\caption{Transmission $T$ versus energy $E$ for para-2 benzene, with $\bar{\beta}=-0.5$, and $\bar{\alpha}=$ (a) $0$ (green solid curve), (b) $-0.2$ (red long-dashed), (c) $-0.4$ (blue short-dashed).}
\label{fig7}
\end{figure}

\begin{figure}[hp]
\includegraphics[width=12cm]{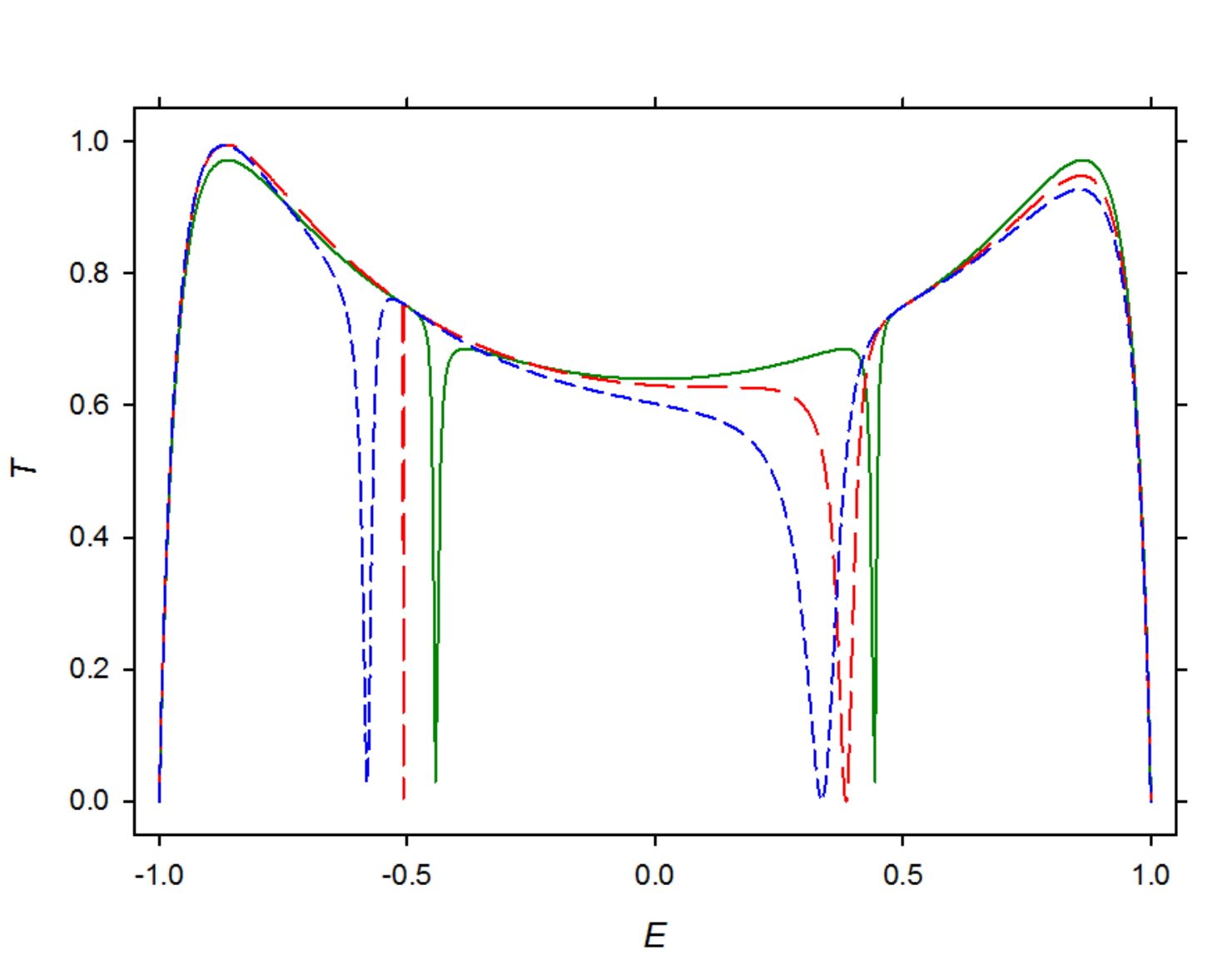}
\caption{Transmission $T$ versus energy $E$ for para-2 benzene, with $\bar{\beta}=-0.4$, and $\bar{\alpha}=$ (a) $0$ (green solid curve), (b) $-0.2$ (red long-dashed), (c) $-0.4$ (blue short-dashed).}
\label{fig8}
\end{figure}

Turning next to para-2 benzene, the curves for  $\bar{\beta}=\beta =-0.5$ are given in Figure \ref{fig7}.
First, note that the curve in Figure \ref{fig7}(a) is again that for pure benzene, which was given above as Figure \ref{fig4}(a).
Now, when $\bar{\alpha} \ne \alpha$ (Figure \ref{fig7}(b,c)), the most striking feature is the creation of a pair of anti-resonances.
Their energies are, approximately, symmetrically placed for smaller $\bar{\alpha}$ (Figure \ref{fig7}(b)), but move to lower
energies as $\bar{\alpha}$ shifts to lower energies (Figure \ref{fig7}(c)).
The positions of the resonance peaks of (a) are relatively unaffected, although their heights are somewhat diminished.
When $\bar{\beta}$ is varied to $-0.4$ (Figure \ref{fig8}), the new anti-resonances persist, and in particular, in the curve (a)
where $\bar{\alpha}=0$, indicating that any perturbation to the pure benzene on site 2 (or an equivalent site) is enough to 
create anti-resonances.
This is an interesting contrast to the para-1 situation, where it remains free of anti-resonances, regardless of the values
of the perturbation parameters. 
Otherwise, the curves of Figure \ref{fig8} look very similar to those of Figure \ref{fig7}(b,c), once again showing the relatively
moderate effect of modifying $\bar{\beta}$.

\begin{figure}[hp]
\includegraphics[width=12cm]{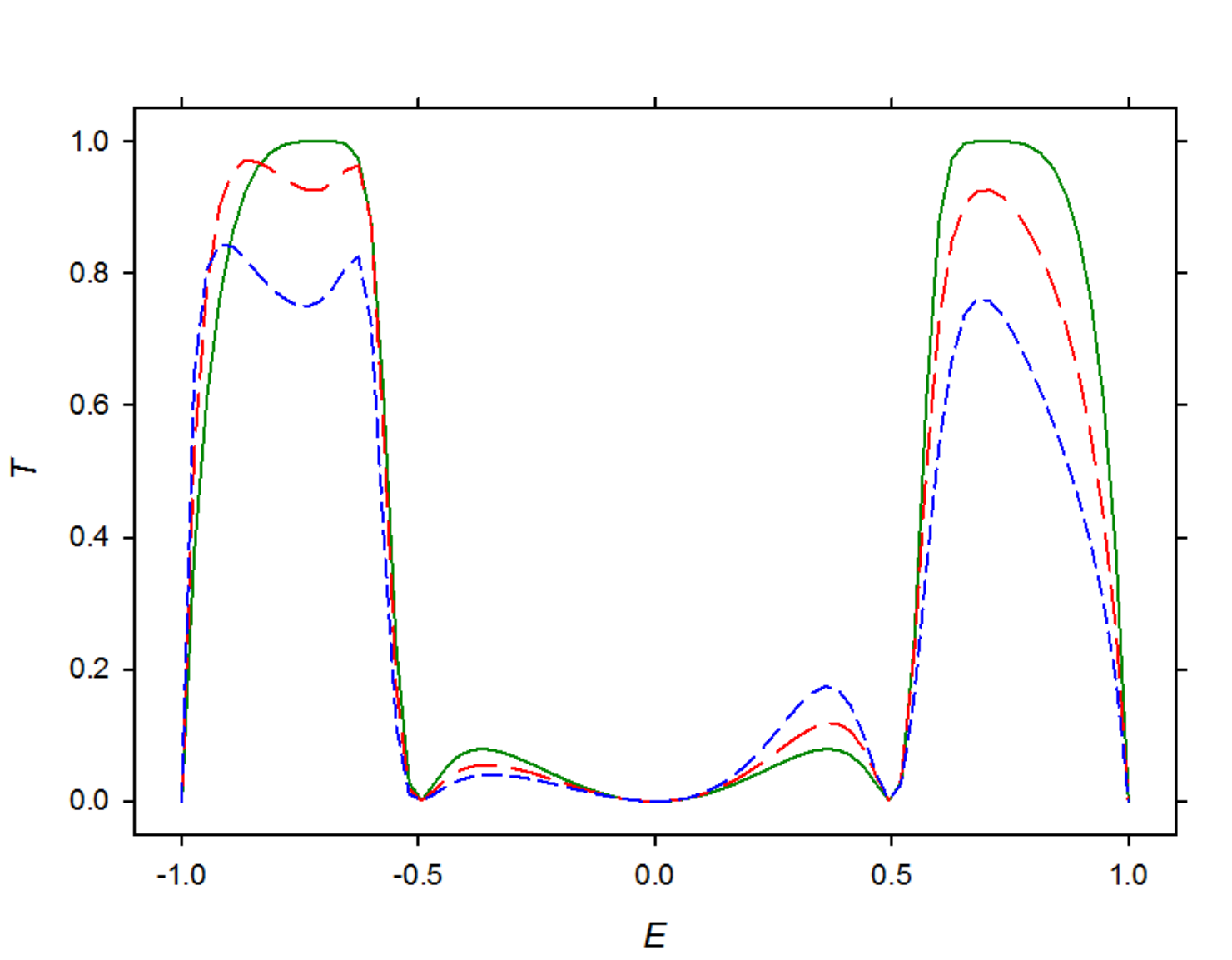}
\caption{Transmission $T$ versus energy $E$ for meta-1 benzene, with $\bar{\beta}=-0.5$, and $\bar{\alpha}=$ (a) $0$ (green solid curve), (b) $-0.2$ (red long-dashed), (c) $-0.4$ (blue short-dashed).}
\label{fig9}
\end{figure}

Next, we consider meta-1 benzene, for which the curves with $\bar{\beta}=\beta =-0.5$ are given in Figure \ref{fig9}.
The graph in Figure \ref{fig9}(a), where $\bar{\alpha}=\alpha=0$ also, is that for pure meta-benzene, from \cite{ref2}.
In this case, there are 3 anti-resonances, at $E=0$ and $\pm 0.5$, which along with the band edges, define four bands
of transmission. 
The two outer bands contain resonances corresponding to strong transmission, while the inner bands have much smaller peaks
and thus weaker transmission. 
On taking $\bar{\alpha} \ne \alpha$ in Figure \ref{fig9}(b,c), it is observed that the anti-resonances are, in fact, pinned,
at the above energies, resulting in comparative stability in the $T(E)$ curves.
Nonetheless, the curves do exhibit some interesting features, such as the fact that the resonances of (a) are lowered
in height as $\bar{\alpha}$ decreases from $\alpha=0$, and the lower resonance shows a splitting into 2 sub-peaks.
Similar observations are made when $\bar{\beta}$ is varied (graphs not shown), namely the anti-resonances are still fixed
in energy, the resonances are reduced in height, and the inner non-resonant peaks show some small increase in height.

\begin{figure}[hp]
\includegraphics[width=12cm]{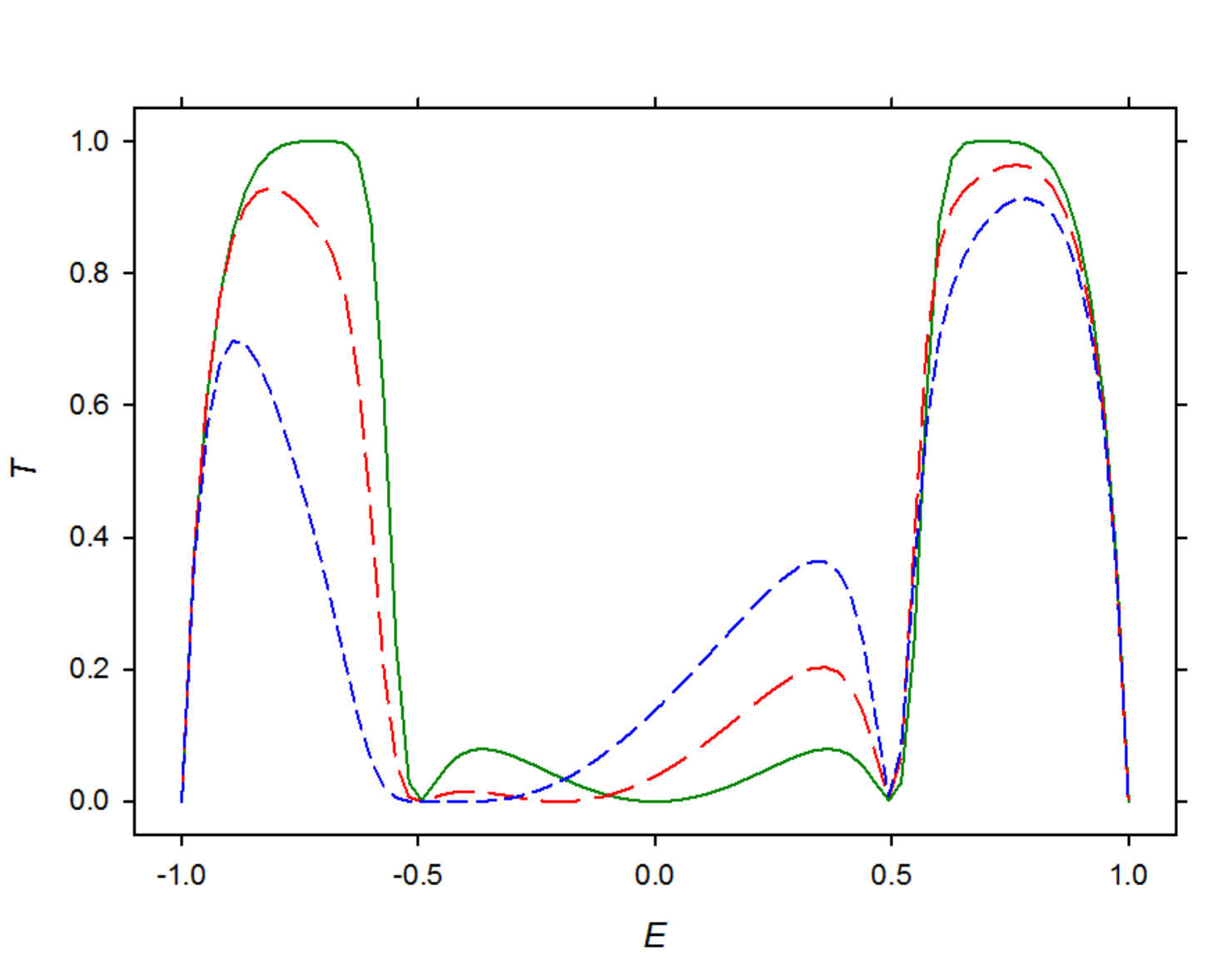}
\caption{Transmission $T$ versus energy $E$ for meta-2 benzene, with $\bar{\beta}=-0.5$, and $\bar{\alpha}=$ (a) $0$ (green solid curve), (b) $-0.2$ (red long-dashed), (c) $-0.4$ (blue short-dashed).}
\label{fig10}
\end{figure}

\begin{figure}[hp]
\includegraphics[width=12cm]{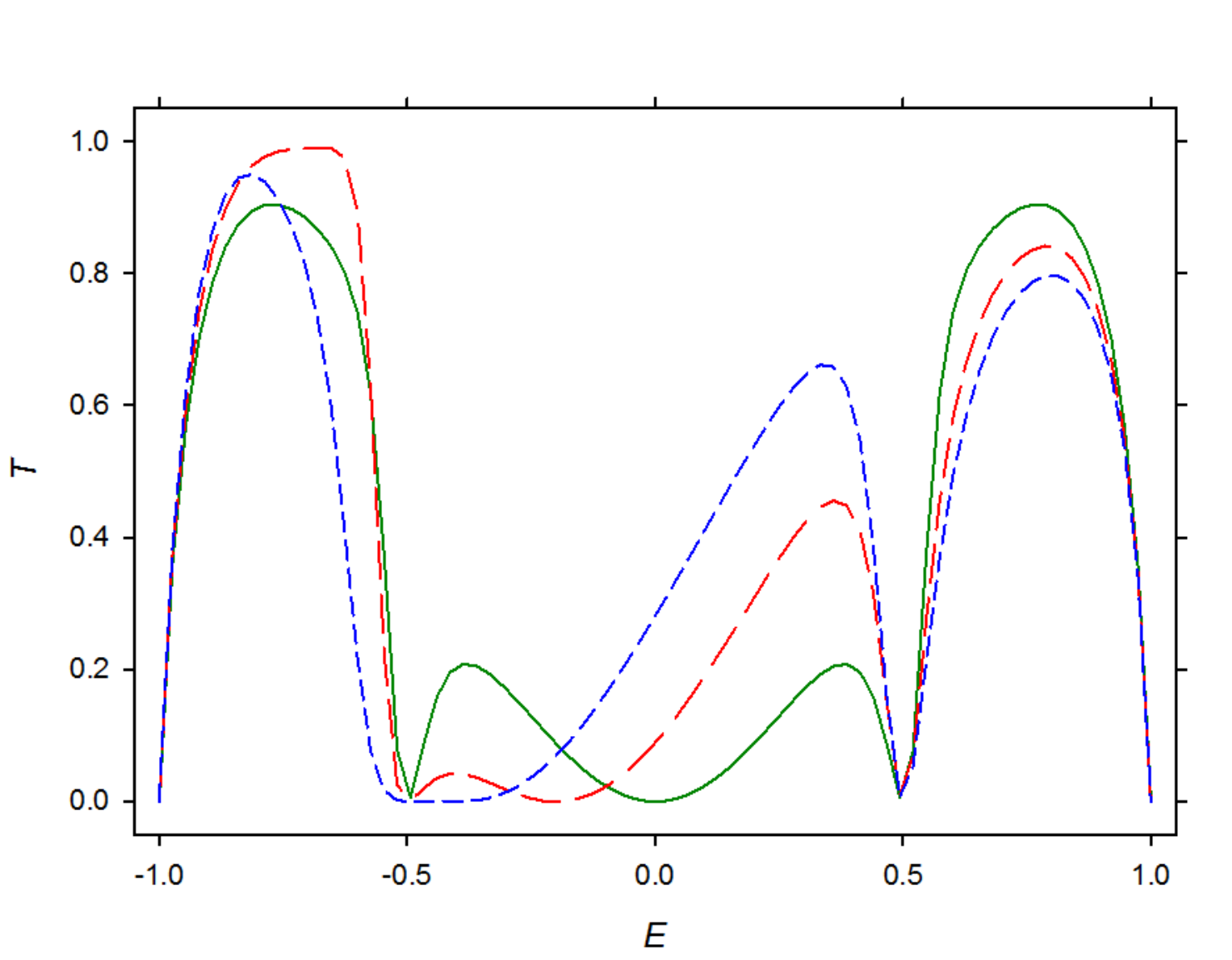}
\caption{Transmission $T$ versus energy $E$ for meta-2 benzene, with $\bar{\beta}=-0.4$, and $\bar{\alpha}=$ (a) $0$ (green solid curve), (b) $-0.2$ (red long-dashed), (c) $-0.4$ (blue short-dashed).}
\label{fig11}
\end{figure}

Moving on, we look at meta-2 benzene, for which the curves with $\bar{\beta}=\beta =-0.5$ are displayed in Figure \ref{fig10}.
The graph for pure benzene is presented again in Figure \ref{fig10}(a), with its 3 anti-resonances, at $E=0$ and $\pm 0.5$.
As $\bar{\alpha}$ is varied, it is seen that the anti-resonances at $E=\pm 0.5$ remain fixed, while that at $E=0$ shifts in energy,
and indeed it is the case that its energy is $E=\bar{\alpha}$.
This is in contrast to the situation for meta-1, where all three anti-resonances remain fixed.
Meanwhile, the two resonance peaks decrease in height, the lower one more substantially than the upper one, while
 the two small inner peaks show some change in height, the upper one increasing quite noticeably, and the lower one
 decreasing almost to the point of disappearance, as the anti-resonance at $E=\bar{\alpha}$ shifts towards coalescence
 with the one at $E=-0.5$.
These features of meta-2 are reinforced when $\bar{\beta}$ is varied from its standard value (Figure \ref{fig11}).
In particular, the positions of the anti-resonances are unchanged from their values in Figure \ref{fig10}, while there
is further rearrangement in the peaks, especially the upper of the 2 inner peaks.

\begin{figure}[hp]
\includegraphics[width=12cm]{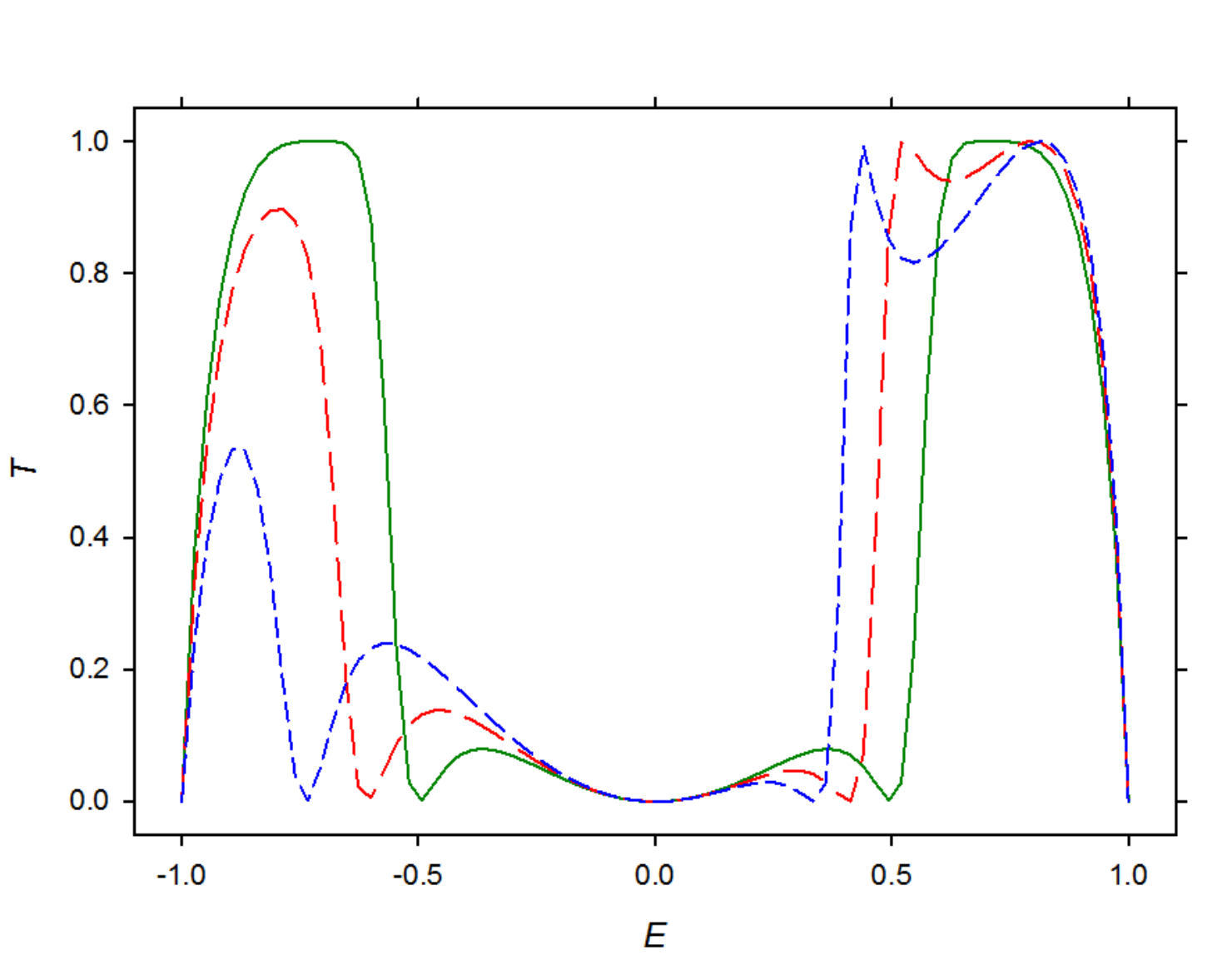}
\caption{Transmission $T$ versus energy $E$ for meta-3 benzene, with $\bar{\beta}=-0.5$, and $\bar{\alpha}=$ (a) $0$ (green solid curve), (b) $-0.2$ (red long-dashed), (c) $-0.4$ (blue short-dashed).}
\label{fig12}
\end{figure}

Now, we turn to meta-3 benzene (Figure \ref{fig12}), with again the pure benzene case given as a reference point (Figure \ref{fig12}(a)).
Varying $\bar{\alpha}$, while keeping $\bar{\beta}=\beta =-0.5$, shows that now it is the anti-resonance at $E=0$ that is pinned,
while the pair at $E=\pm 0.5$ shift to lower energies (as $\bar{\alpha}$ does so). 
This is in striking contrast to the situations seen above for meta-1 and meta-2.
Simultaneously, there is rearrangement in the peaks, most noticeably a diminishment of the lowest (originally-resonant) peak
and a partial splitting of the uppermost one.
Furthermore (graphs not shown), the two anti-resonances at $E=\pm 0.5$ do in fact shift with variation of $\bar{\beta}$
(in contrast to meta-1 and meta-2), moving closer to each other and to the pinned anti-resonance at $E=0$.
The consequent effect, then, is to further diminish the inner pair of peaks while amplifying the outer pair.

\begin{figure}[hp]
\includegraphics[width=12cm]{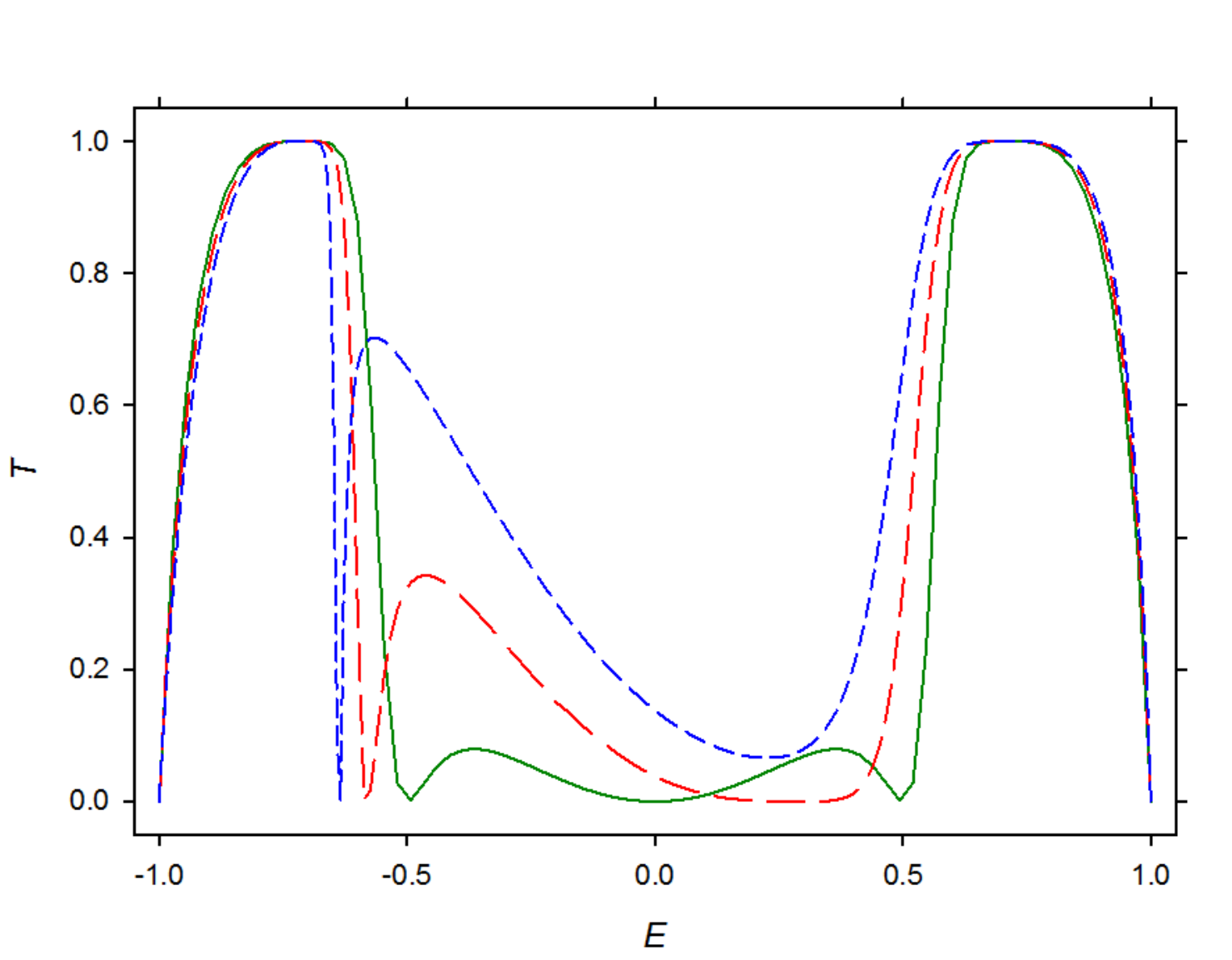}
\caption{Transmission $T$ versus energy $E$ for meta-6 benzene, with $\bar{\beta}=-0.5$, and $\bar{\alpha}=$ (a) $0$ (green solid curve), (b) $-0.2$ (red long-dashed), (c) $-0.4$ (blue short-dashed).}
\label{fig13}
\end{figure}

Lastly in the meta-family, we come to meta-6 benzene, with pure benzene shown again, for reference, in Figure \ref{fig13}(a).
As $\bar{\alpha}$ is varied, something interesting and unique to this system occurs (Figure \ref{fig13}(b,c)). 
The lowest anti-resonance, at $E=-0.5$, merely shifts, rather mundanely, to somewhat lower energies.
However, the other two anti-resonances, at $E=0$ and $E=0.5$ shift closer to one another (Figure \ref{fig13}(b)),
and eventually (for $\bar{\alpha} \approx 0.23$) coalesce and then disappear (Figure \ref{fig13}(c)), leaving the lowest
anti-resonance as the sole survivor.
This remarkable occurrence has substantial impact on the $T(E)$ curve, specifically by now allowing significant 
transmission in the mid-band energy regime.
When $\bar{\beta}$ is varied (graphs not shown), this effect can even be strengthened, although the structure
of the $T(E)$ curve is dominated by the number and positions of the anti-resonances.

\begin{figure}[hp]
\includegraphics[width=12cm]{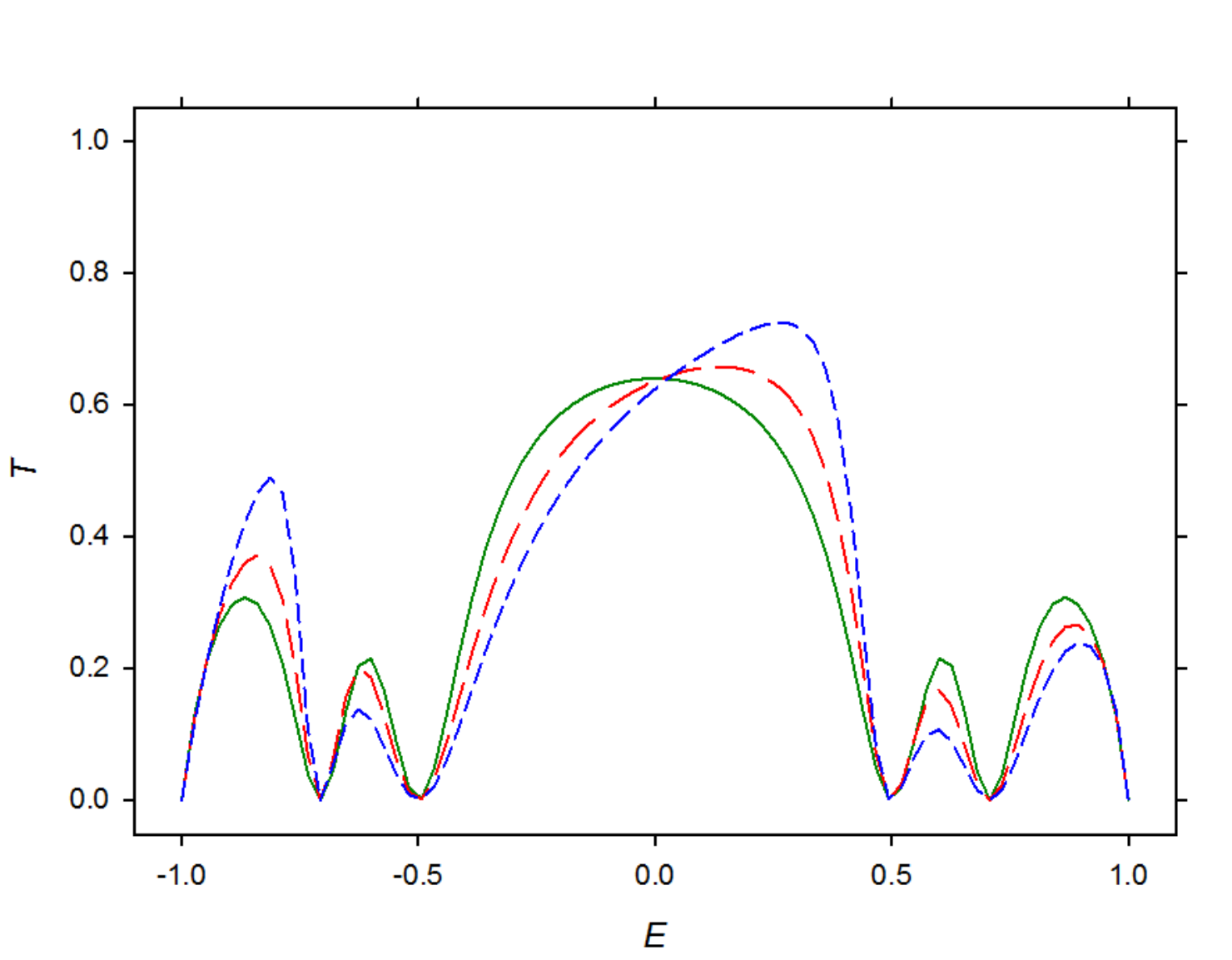}
\caption{Transmission $T$ versus energy $E$ for ortho-1 benzene, with $\bar{\beta}=-0.5$, and $\bar{\alpha}=$ (a) $0$ (green solid curve), (b) $-0.2$ (red long-dashed), (c) $-0.4$ (blue short-dashed).}
\label{fig14}
\end{figure}

We now turn our attention to the ortho-benzene group,beginning with ortho-1. 
The starting point is, once again, the pure benzene case, which is shown in Figure \ref{fig14}(a) \cite{ref2}.
It can be seen that there are 4 anti-resonances, at $E= \pm 0.5$ and $E= \pm \sqrt2 /2$, thus dividing the band into 5
regions of transmission (none resonant), with the middle one clearly being dominant.
On varying $\bar{\alpha}$, while keeping $\bar{\beta}$ constant (Figure \ref{fig14}(b,c)), it is observed that all 4 anti-resonances
remain pinned, and there is some moderate rearrangement within the regions of transmission, with some increasing in height
and others decreasing.
Upon allowing $\bar{\beta}$ to vary as well (graphs not shown), the anti-resonances remain in their original positions,
with the result that all graphs look very much the same.
These features are qualitatively similar to those seen in the para-1 and meta-1 situations, namely graphs dominated by
anti-resonances that are fixed in number and position.

\begin{figure}[hp]
\includegraphics[width=12cm]{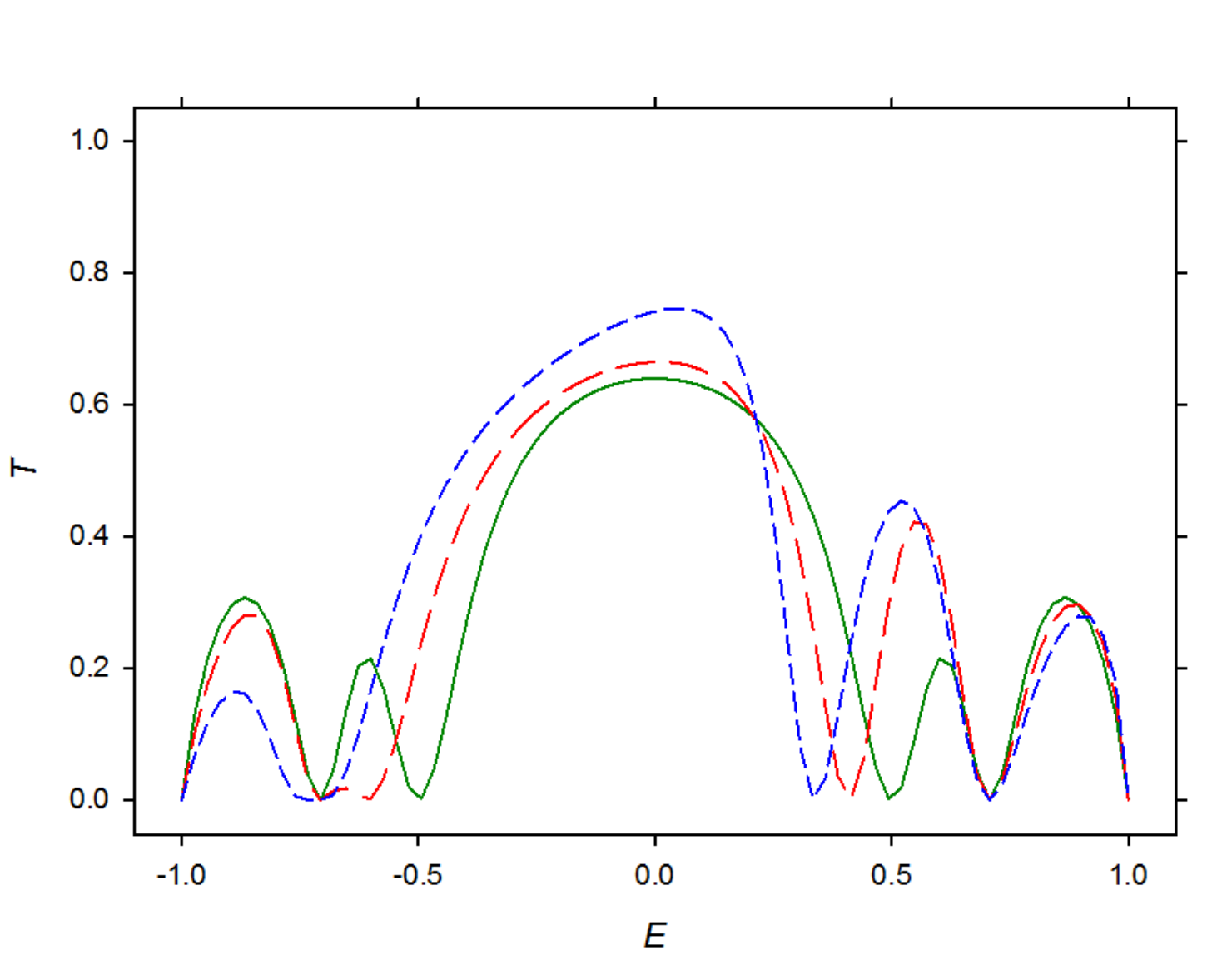}
\caption{Transmission $T$ versus energy $E$ for ortho-2 benzene, with $\bar{\beta}=-0.5$, and $\bar{\alpha}=$ (a) $0$ (green solid curve), (b) $-0.2$ (red long-dashed), (c) $-0.4$ (blue short-dashed).}
\label{fig15}
\end{figure}

Next, we look at ortho-2 benzene (Figure \ref{fig15}), with Figure \ref{fig15}(a) showing the pure case.
Variation of $\bar{\alpha}$, as seen in Figure \ref{fig15}(b,c), indicates that the outer pair of anti-resonances, at
$E= \pm \sqrt2 /2$, are fixed in energy, while the inner pair, at $E= \pm 0.5$ are not. 
This situation allows for considerable rearrangement of the $T(E)$ curve and its sub-regions, as is clear from
the graphs.
Indeed, in Figure \ref{fig15}(c), it can be seen that the anti-resonance, originally at $E=-0.5$, is shifted  downwards
in energy, far enough so that it coalesces with the one at $E= -\sqrt2 /2$.
Variation of $\bar{\beta}$ (graphs not shown) still leaves the outer anti-resonances pinned at $E= \pm \sqrt2 /2$, 
while allowing the inner pair to move even further, so that transmission profiles significantly different from the pure case
can be achieved.

\begin{figure}[hp]
\includegraphics[width=12cm]{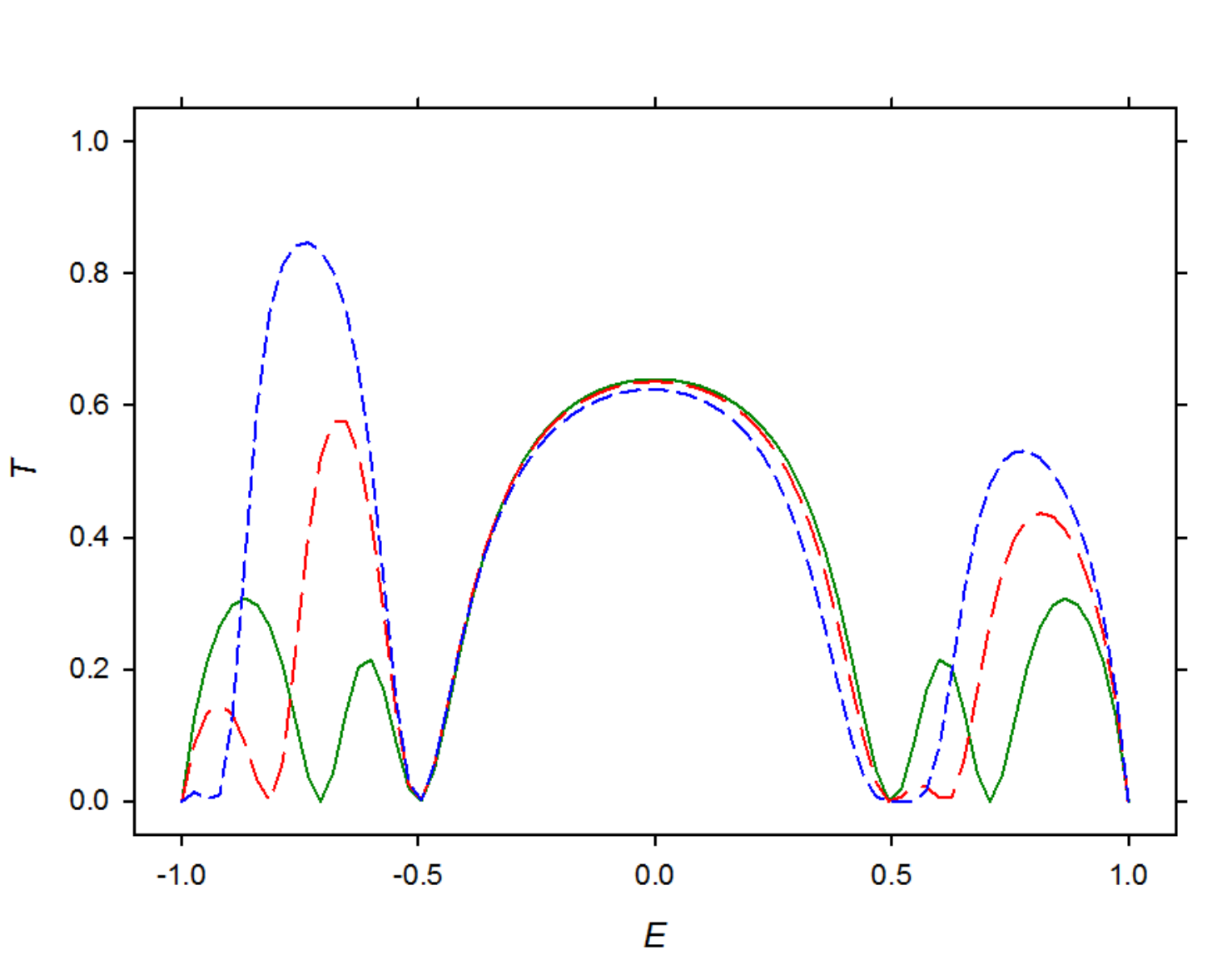}
\caption{Transmission $T$ versus energy $E$ for ortho-3 benzene, with $\bar{\beta}=-0.5$, and $\bar{\alpha}=$ (a) $0$ (green solid curve), (b) $-0.2$ (red long-dashed), (c) $-0.4$ (blue short-dashed).}
\label{fig16}
\end{figure}

Lastly, we arrive at ortho-3 benzene, with (as usual) the pure case shown in Figure \ref{fig16}(a).
On varying $\bar{\alpha}$ (with $\bar{\beta}=\beta =-0.5$), we can see that, in contrast to ortho-1 and ortho-2,
it is now the inner pair of anti-resonances (at $E= \pm 0.5$)  that are pinned, while the outer pair (initially at $E= \pm \sqrt2 /2$)
move to lower energies as $\bar{\alpha}$ does.
As a result, the dominant central peak is relatively unaffected by changes in $\bar{\alpha}$, while the outer regions show
huge variations, even to the extent that two of these regions all but disappear in Figure \ref{fig16}(c).
This behaviour persists when $\bar{\beta}$ is allowed to vary (graphs not shown), because,
although the anti-resonances at $E= \pm 0.5$ still remain fixed, again the movement of the other pair allows
considerable rearrangement of the transmission curve, and indeed significant enhancement of transmission
near the band edges.

\section{Conclusions}

In this paper, we have used the renormalization method to study the electron transmission properties of a benzene molecule, in which one of its
carbon atoms has been perturbed or replaced.
The different combinations of lead attachments and location of the modified atom produce a total of 9
variations on the basic system. 
Consequently, a rich variety of behaviours are observed in the $T(E)$ curves, as the perturbation parameters
$\bar{\alpha}$ and $\bar{\beta}$ are varied.
As would be expected, the basic shape of the $T(E)$ curve is primarily determined by the number and locations
of anti-resonances and maxima (resonant or not).
Of the two parameters, the site energy $\bar{\alpha}$ is seen to be the more important, with the effect of
moderate changes in $\bar{\beta}$ usually being fairly modest.
A general observation is that the $T(E)$ curve is symmetric whenever $\bar{\alpha}=\alpha$, regardless
of whether $\bar{\beta}$ equals $\beta$ or not.
The clearest cases are those for which $l=1$, namely when the modified atom is attached directly to one
of the leads.
In these situations (para-1, meta-1, ortho-1), the number of anti-resonances is unchanged by variation
of the parameters, and those that do exist are pinned in energy.
As a result, changes in the $T(E)$ curve with $\bar{\alpha}$ and/or $\bar{\beta}$ are rather limited, so these
systems are fairly close to the corresponding ones involving pure benzene.
More diverse, and thus more interesting, behaviour arises when $l \ne 1$, wherein some or all of the anti-resonances
shift in energy as the parameters are varied, and moreover, there are cases where new anti-resonances
are created (para-2) or existing ones disappear (meta-6).
In these cases, the anti-resonances that are pinned remain so for all parameter values, while those that
are not pinned can shift in energy with either $\bar{\alpha}$ or $\bar{\beta}$ or both.
Thus, the $l \ne 1$ systems show more promise from the perspective of circuit design, in that they
allow for much greater flexibility in fine-tuning the desired transmission properties, and in some cases
(such as meta-2), significantly enhance the transmittivity.

\section{Acknowledgment}

We thank Prof. Sydney G. Davison for helpful discussions.

\section{Keywords}

electron transmission, benzene, renormalization method

\end{document}